\documentclass[a4,12pt]{article}
\usepackage[small,compact]{titlesec}
  
\usepackage[latin1]{inputenc}
\usepackage[T1]{fontenc}
\usepackage{rotating}
\usepackage{multirow}
\usepackage{array}
\usepackage{graphicx}
\usepackage{epstopdf}
 
 \usepackage{amsbsy}
\usepackage{amssymb}
\usepackage{amscd}
\usepackage{amsmath}

 \usepackage{times,subfigure,epsfig}
 
\usepackage{cite}
\usepackage{latexsym}
\usepackage{graphicx}
\usepackage{amsmath}
\usepackage{amssymb}
\usepackage{amsfonts}
\usepackage{psfrag}
 
\usepackage[small,bf]{caption}

\newlength{\figwidth}
\newlength{\ffh}
\newlength{\ffw}
\begin{document}

\setlength{\unitlength}{1cm}

\pagestyle{myheadings}

\setcounter{page}{1}

\begin{center}
\large \textsc{Massive MIMO for Next Generation Wireless Systems} \\~\\
\normalsize Erik G. Larsson, ISY, Link\"oping University, Sweden\\
\normalsize Ove Edfors, Lund  University, Sweden \\
\normalsize Fredrik Tufvesson, Lund  University, Sweden \\
\normalsize Thomas L. Marzetta, Bell Labs, Alcatel-Lucent, USA

\today
\end{center}

\begin{abstract}
Multi-user Multiple-Input Multiple-Output (MIMO) offers big advantages over conventional point-to-point
MIMO: it works with cheap single-antenna terminals, a rich scattering
environment is not required, and resource allocation is simplified
because every active terminal utilizes all of the time-frequency
bins. However, multi-user MIMO, as originally envisioned with roughly
equal numbers of service-antennas and terminals and frequency division duplex operation, is
not a scalable technology. 

\emph{Massive MIMO} (also known as ``Large-Scale Antenna Systems'',
``Very Large MIMO'', ``Hyper MIMO'', ``Full-Dimension MIMO''
and ``ARGOS'') makes a clean break with current
practice through the use of a large excess of service-antennas over
active terminals and time division duplex operation. Extra antennas help by focusing
energy into ever-smaller regions of space to bring huge improvements
in throughput and radiated energy efficiency. Other benefits of
massive MIMO include the extensive use of inexpensive low-power
components, reduced latency, simplification of the media access control (MAC) layer, and
robustness to intentional jamming. The anticipated throughput depend
on the propagation environment providing asymptotically orthogonal
channels to the terminals, but so far experiments have not disclosed
any limitations in this regard. While massive MIMO renders many
traditional research problems irrelevant, it uncovers entirely new
problems that urgently need attention: the challenge of making many
low-cost low-precision components that work effectively together,
acquisition and synchronization for newly-joined terminals, the
exploitation of extra degrees of freedom provided by the excess of
service-antennas, reducing internal power consumption to achieve total
energy efficiency reductions, and finding new deployment scenarios.
This paper presents an overview of the massive MIMO concept and
of contemporary research on the topic.
\end{abstract}

\section{Background: Multi-User MIMO Maturing}
 
\emph{MIMO, Multiple-Input Multiple Output, technology} relies on multiple antennas to
simultaneously transmit multiple streams of data in wireless
communication systems. When MIMO is used to communicate with several
terminals at the same time, we speak of \emph{multiuser MIMO}. Here,
we just say MU-MIMO for short.

MU-MIMO in cellular systems brings improvements on four fronts:
\begin{itemize}
\item \emph{increased data rate}, because the more antennas, the more
  independent data streams can be sent out and the more terminals can
  be served simultaneously;

\item \emph{enhanced reliability}, because the more antennas the more
  distinct paths that the radio signal can propagate over;

\item \emph{improved energy efficiency}, because the base station can
  focus its emitted energy into the spatial directions where it knows
  that the terminals are located; and

\item \emph{reduced interference} because the base station can
  purposely avoid transmitting into directions where spreading
  interference would be harmful.
\end{itemize}
All improvements cannot be achieved simultaneously, and there are
requirements on the propagation conditions, but the four above bullets
are the general benefits.  MU-MIMO technology for wireless communications
in its conventional form is maturing, and incorporated into recent and
evolving wireless broadband standards like 4G LTE and LTE-Advanced
(LTE-A). The more antennas the base station (or terminals) are
equipped with, the better performance in all the above four respects---at least for operation in time-division duplexing (TDD) mode.
However, the number of antennas used today is modest. The most modern
standard, LTE-Advanced, allows for up to 8 antenna ports at the base
station and equipment being built today has much fewer antennas than
that.

\section{Going Large: Massive MIMO} 

\underline{Massive MIMO} is an emerging technology, that scales up
MIMO by possibly orders of magnitude compared to current
state-of-the-art.  In this paper, we follow up on our earlier
exposition \cite{RPL2013}, with a focus on the developments in the
last three years: most particularly, energy efficiency, exploitation
of excess degrees of freedom, TDD calibration, techniques to
combat pilot contamination, and entirely new channel measurements.

\begin{figure}[t!]
 \centerline{ \includegraphics[width=10cm] {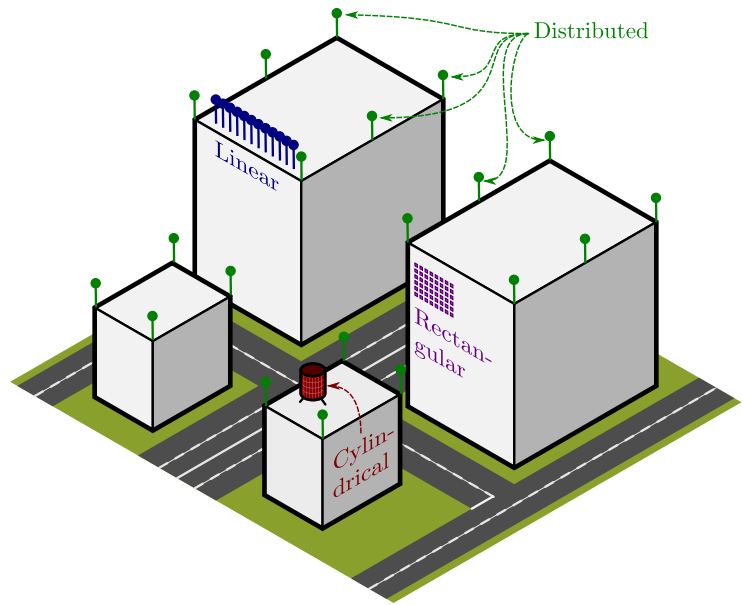}
 } \caption{\label{fig:new_deploy}Some possible antenna configurations and
   deployment scenarios for a massive MIMO base station.}
 \end{figure}

With massive MIMO, we think of systems that use antenna arrays with a
few hundred antennas, simultaneously serving many tens of terminals in
the same time-frequency resource.  The basic premise behind massive
MIMO is to reap all the benefits of conventional MIMO, but on a much
greater scale.  Overall, massive MIMO is an enabler for the
development of future broadband (fixed and mobile) networks which will
be energy-efficient, secure, and robust, and will use the spectrum
efficiently.  As such, it is an enabler for the future digital society
infrastructure that will connect the Internet of people, Internet of
things, with clouds and other network infrastructure.  Many different
configurations and deployment scenarios for the actual antenna arrays
used by a massive MIMO system can be envisioned, see
Fig.~\ref{fig:new_deploy}. Each antenna unit would be small, and
active, preferably fed via an optical or electric digital bus.

 Massive MIMO relies on spatial multiplexing that in turn relies on
 the base station having good enough channel knowledge, both on the
 uplink and the downlink.  On the uplink, this is easy to accomplish
 by having the terminals send pilots, based on which the base station
 estimates the channel responses to each of the terminals. The
 downlink is more difficult. In conventional MIMO systems, like the
 LTE standard, the base station sends out pilot waveforms based on
 which the terminals estimate the channel responses, quantize the
 so-obtained estimates and feed them back to the base station. This
 will not be feasible in massive MIMO systems, at least not when
 operating in a high-mobility environment, for two reasons.  First,
 optimal downlink pilots should be mutually orthogonal between the
 antennas. This means that the amount of time-frequency resources
 needed for downlink pilots scales as the number of antennas, so a
 massive MIMO system would require up to a hundred times more such
 resources than a conventional system. Second, the number of channel
 responses that each terminal must estimate is also proportional to
 the number of base station antennas. Hence, the uplink resources
 needed to inform the base station about the channel responses would
 be up to a hundred times larger than in conventional systems.  Generally, the
 solution is to operate in TDD mode, and rely
 on reciprocity between the uplink and downlink channels---although FDD operation may be possible
 in certain cases \cite{nam2012joint}.
 
While the concepts of massive MIMO have been mostly theoretical so
far, and in particular stimulated much research in random matrix
theory and related mathematics, basic testbeds are becoming available
\cite{SYA2012} and initial channel measurements have been performed
\cite{GTER2012,HHW2012}.

\section{The Potential of Massive MIMO}\label{sec:potential}

Massive MIMO technology relies on phase-coherent but computationally
very simple processing of signals from all the antennas at the
base station.  Some specific benefits of a massive MU-MIMO system are:
\begin{itemize}
\item Massive MIMO can \emph{increase the capacity} 10 times or more
  and \emph{simultaneously}, \emph{improve the radiated energy-efficiency} in
  the order of 100 times.

The \emph{capacity increase} results from the aggressive spatial multiplexing
used in massive MIMO.  The fundamental principle that makes the
dramatic increase in \emph{energy efficiency} possible is that with large
number of antennas, energy can be focused with extreme sharpness into small
regions in space, see Fig.~\ref{focus}. The underlying physics is
\emph{coherent superposition} of wavefronts. By appropriately shaping
the signals sent out by the antennas, the base station can make sure
that all wave fronts collectively emitted by all antennas add up
\emph{constructively} at the locations of the intended terminals, but
\emph{destructively (randomly)} almost everywhere else. Interference
between terminals can be suppressed even further by using, \emph{e.g.},
zero-forcing (ZF). This, however, may come at the cost of more transmitted power,
as illustrated in Fig.~\ref{focus}.

\begin{figure}[t!]
	\centerline{
		\includegraphics[width=16 cm]{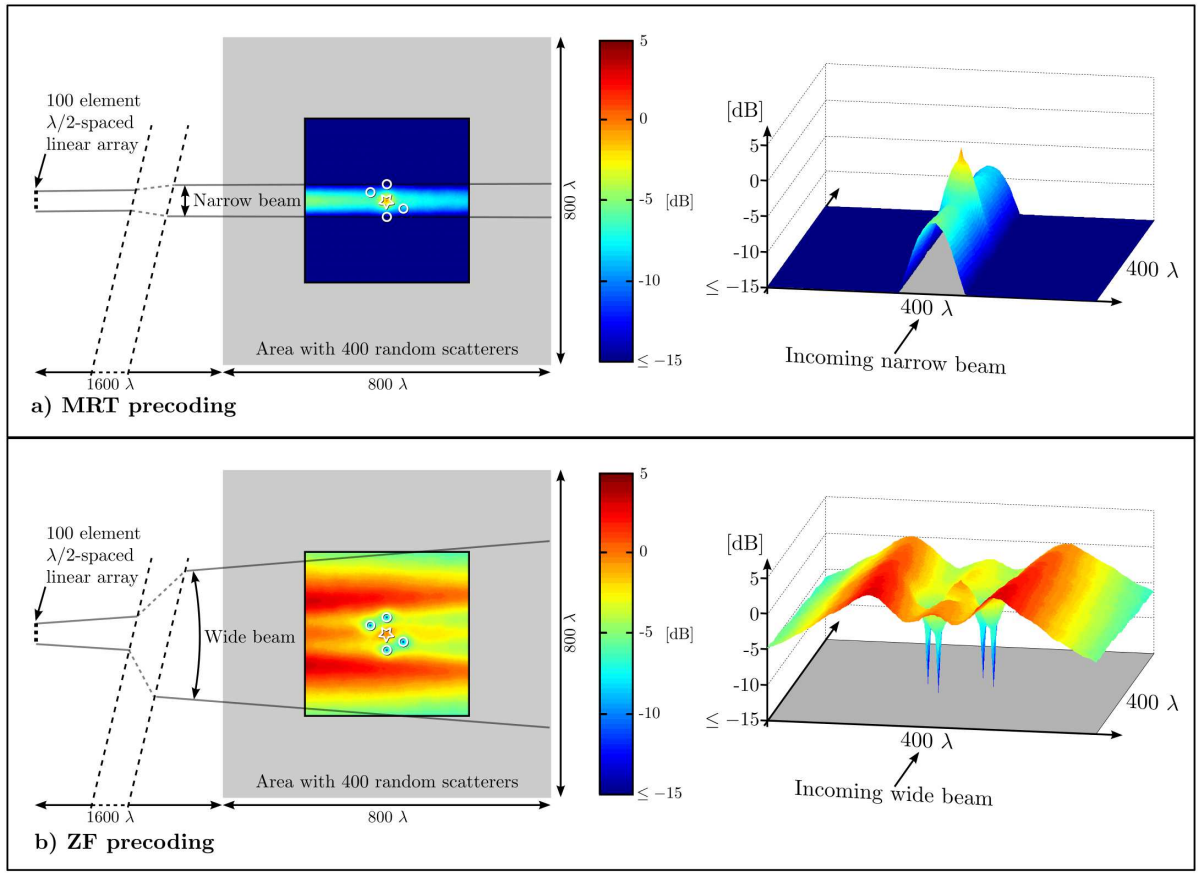}
	}
	\caption{Relative field strength around a target terminal in a
          scattering environment of size $800\lambda \times 800
          \lambda$, when the base station is placed $1600 \lambda$ to
          the left. Average field strengths are calculated over
          10000 random placements of 400 scatterers, when two different
          linear precoders are used: a) MRT
          precoders and b) ZF precoders. \emph{Left:} pseudo-color plots of average field strengths,
          with target user positions at the center ($\star$), and four other
          users nearby ($\circ$). \emph{Right:} average field
          strengths as surface plots, allowing an alternate view of
          the spatial focusing.\label{focus}}
\end{figure}

More quantitatively, Fig.~\ref{mrc} (from \cite{NLM2013}) depicts the fundamental tradeoff
between the \emph{energy efficiency} in terms of the total number of bits (sum-rate) transmitted per
Joule per terminal receiving service of energy spent, and \emph{spectral efficiency} in terms of total number of bits (sum-rate)
transmitted per unit of radio spectrum consumed.  The figure illustrates
the relation for the uplink, from the terminals to the base station (the downlink performance is similar).
The figure shows the tradeoff for three cases:
\begin{figure}[t!]
\centerline{
\begin{psfrags}
          \psfrag{u}[cu]{$ 2$}
 \includegraphics[width=11cm]
      {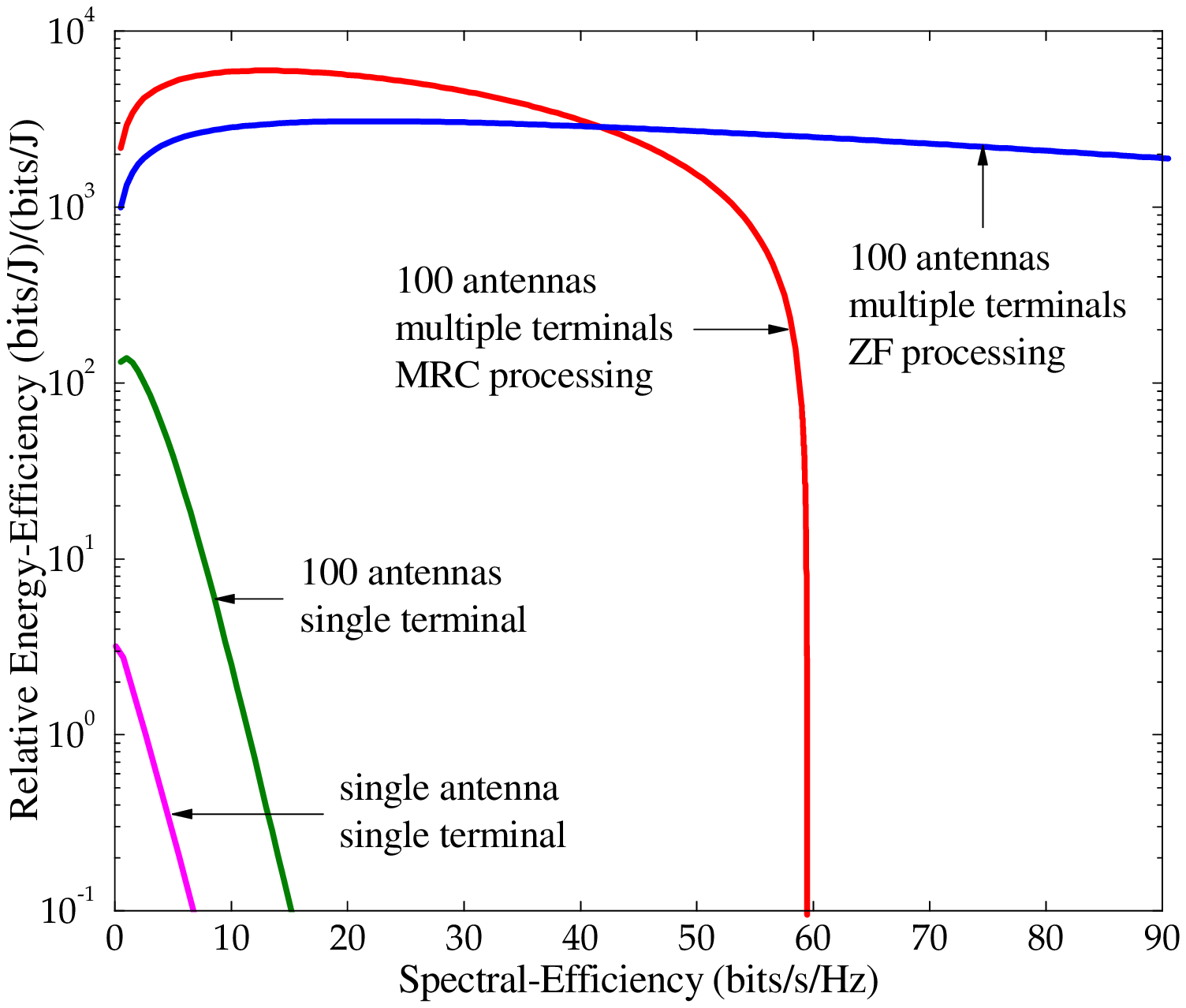}
\end{psfrags}
} 
 \caption{Half the power---twice the force (from \cite{NLM2013}): Improving uplink spectral
   efficiency 10 times and simultaneously increasing the
   radiated-power efficiency 100 times with massive MIMO technology,
   using extremely simple signal processing---taking into account the energy and bandwidth costs of
   obtaining channel state information.  \label{mrc}}
\end{figure}
\begin{itemize}
\item  a reference system with one single antenna serving a single terminal (purple),

\item a system with 100 antennas serving a single terminal using
  conventional beamforming (green)

\item a massive MIMO system with 100 antennas simultaneously serving
  multiple (about 40 here) terminals (red, using maximum-ratio combining; and blue, using zero-forcing).
\end{itemize}
The attractiveness of maximum-ratio 
combining (MRC) compared with
ZF is not only its computational
simplicity---multiplication of the received signals by the conjugate
channel responses, but also that it can be performed in a distributed
fashion, independently at each antenna unit.  While ZF also works
fairly well for a conventional or moderately-sized MIMO system, MRC
generally does not. The reason for why MRC works so well for massive
MIMO is that the channel responses associated with different terminals
tend to be nearly orthogonal when the number of base station antennas
is large---see Section~\ref{sec:prop}.  

The prediction in
Fig.~\ref{mrc} is based on an information-theoretic analysis that
takes into account intracell interference, as well as the bandwidth
and energy cost of using pilots to acquire channel state information
in a high-mobility environment \cite{NLM2013}.  With the MRC receiver, we operate in the nearly
noise-limited regime of information theory. This means providing each
terminal with a rate of about 1 bit per complex dimension (1
bps/Hz). In a massive MIMO system, when using MRC and when operating
in the ``green'' regime, that is, scaling down the power as much as
possible without seriously affecting the overall spectral efficiency,
multiuser interference and effects from hardware imperfections tend to
be overwhelmed by the thermal noise.  The reason that the overall spectral
efficiency still can be 10 times higher than in conventional MIMO is
that many tens of terminals are served simultaneously, \emph{in the
  same time-frequency resource}.  When operating in the 1
bit/dimension/terminal regime, there is also some evidence that
intersymbol interference can be treated as additional thermal noise
\cite{PML2012}, hence offering a way of disposing with OFDM as a means
of combatting intersymbol interference.

To understand the scale of the capacity gains that massive MIMO
offers, consider an array consisting of 6400 omnidirectional antennas
(total form factor $6400\times (\lambda/2)^2 \approx 40$ m$^2$),
transmitting with a total power of 120 Watts (that is, each antenna
radiating about 20 mW) over a 20 MHz bandwidth in the PCS band (1900
MHz). The array serves one thousand (1000) fixed terminals randomly
distributed in a disk of radius 6 km centered on the array, each
terminal having an 8 dB gain antenna. The height of the antenna array
is 30 m, and the height of the terminals is 5 m. Using the
Hata-COST231 model we find that the path loss is 127 dB at 1 km range
and the range-decay exponent is 3.52. There is also log-normal shadow
fading with 8 dB standard deviation. The receivers have a 9 dB noise
figure. One-quarter of the time is spent on transmission of uplink
pilots for TDD channel estimation, and it is assumed that the channel
is substantially constant over intervals of 164 ms in order to
estimate the channel gains with sufficient accuracy.  Downlink data is
transmitted via maximum-ratio transmission (MRT) beamforming combined
with power control, where the 5\% of the terminals having the worst
channels are excluded from service.  We use a capacity lower bound
from \cite{yang2013performance} that is extended to accommodate slow
fading, near/far effects and power control and which accounts for
receiver noise, channel estimation errors, the overhead of pilot
transmission, and the imperfections of MRT beamforming. We use optimal
max-min power control which confers an equal
signal-to-interference-and-noise ratio on each of the 950 terminals
and therefore equal throughput. Numerical averaging over random
terminal locations and over the shadow fading shows that 95\% of the
terminals will receive a throughput of 21.2 Mb/s/terminal.  Overall,
the array in this example will offer the 1000 terminals a total
downlink throughput of 20 Gb/s, resulting in a sum-spectral efficiency
of 1000 bits/s/Hz. This would be enough, for example, to provide 20
Mbit/s broadband service to each of a thousand homes. The max-min
power control provides equal service \emph{simultaneously} to 950
terminals. Other types of power control combined with time-division
multiplexing could accommodate heterogeneous traffic demands of a
larger set of terminals.

The MRC receiver (for the uplink) and its counterpart MRT precoding
(for the downlink) are also known as matched filtering (MF) in the literature.

\item Massive MIMO can be built with \emph{inexpensive, low-power components.}

  Massive MIMO is a game-changing technology both with regard to
  theory, systems and implementation.  With massive MIMO, expensive,
  ultra-linear 50 Watt amplifiers used in conventional systems are
  replaced by hundreds of low-cost amplifiers with output power in the
  milli-Watt range.  The contrast to classical array designs, which
  use few antennas fed from  high-power amplifiers, is significant.
  Several expensive and bulky items, such as large coaxial cables, can
  be eliminated altogether. (The typical coaxial cables used for
  tower-mounted base stations today are more than four centimeters in
  diameter!)

  Massive MIMO reduces the constraints on accuracy and linearity of
  each individual amplifier and RF chain. All what matters is their
  combined action.  In a way, massive MIMO relies on the law of
  large numbers to make sure that noise, fading and hardware
  imperfections average out when signals from a large number of
  antennas are combined in the air together.  The same property that
  makes massive MIMO resilient against fading also makes the
  technology extremely robust to failure of one or a few of the
  antenna units.

  A massive MIMO system has a large surplus of degrees of freedom. For
  example, with 200 antennas serving 20 terminals, 180 degrees of
  freedom are unused. These degrees of freedom can be used for
  hardware-friendly signal shaping.  In particular, each antenna can
  transmit signals with very small peak-to-average ratio
  \cite{StL2013} or even constant envelope \cite{MoL2013} at a very
  modest penalty in terms of increased total radiated power.  Such
  (near-constant) envelope signaling facilitates the use of extremely
  cheap and power-efficient RF amplifiers.  The techniques in
  \cite{StL2013,MoL2013} must not be confused with conventional
  beamforming techniques or equal-magnitude-weight beamforming
  techniques.  This distinction is explained in Fig.~\ref{ce}. With
  (near) constant-envelope multiuser precoding, no beams are formed,
  and the signals emitted by each antenna are not formed by weighing of
  a symbol. Rather,  a wavefield is created, such that when this
  wavefield is sampled at the spots where the terminals are located,
  the terminals see precisely the signals that we want them to see.
  The fundamental property of the massive MIMO channel that makes this
  possible is that the channel has a large nullspace: almost anything
  can be put into this nullspace without affecting what the terminals
   see. In particular, components can be put into this nullspace
  that make the transmitted waveforms satisfy the desired envelope constraints.
   Notwithstanding that, the \emph{effective channels}
  between the base station and each of the terminals, can take any
  signal constellation as input and does \emph{not} require the use of
  PSK-type modulation.

\begin{figure}[t!]
\centerline{
 \includegraphics[width=16cm]
      {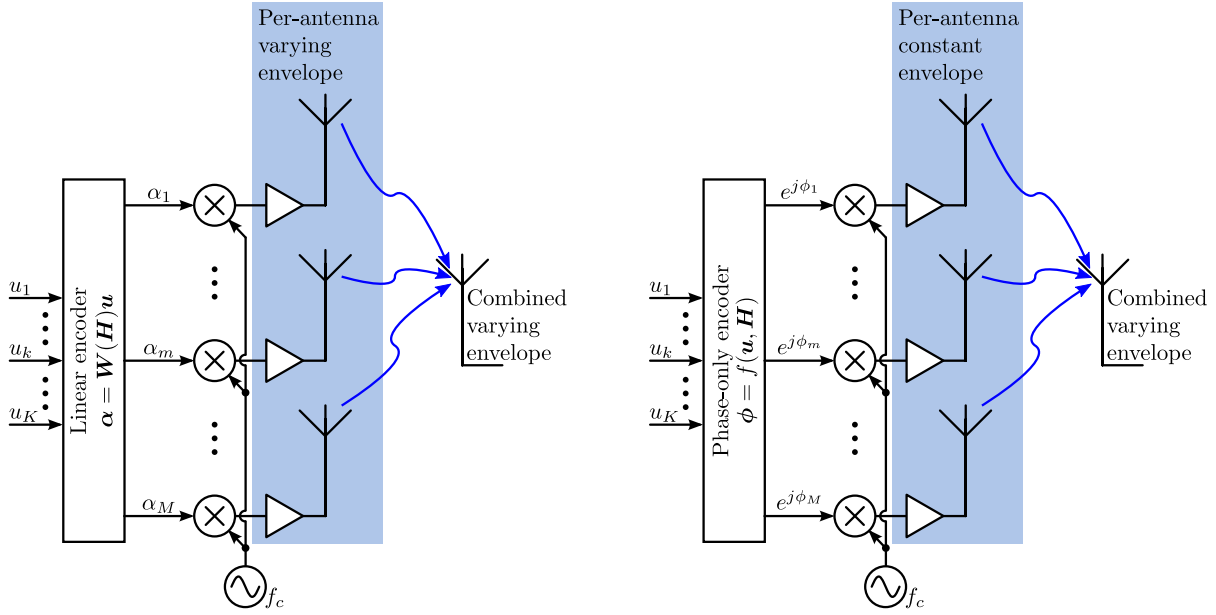}
}
\caption{\label{ce}Conventional MIMO beamforming, contrasted with
  per-antenna constant-envelope transmission in massive MIMO. \emph{Left:}
  conventional beamforming, where the signal emitted by each antenna has a
  large dynamic range. \emph{Right:} per-antenna constant-envelope
  transmission, where each antenna sends out a signal with constant
  envelope.}
\end{figure}

The drastically improved energy efficiency enables massive MIMO
systems to operate with a total output RF power two orders of magnitude
less than with current technology.  This matters, because the energy
consumption of cellular base stations is a growing concern worldwide.
In addition, base stations that consume many orders of magnitude less
power could be powered by wind or solar, and hence easily deployed
where no electricity grid is available.  As a bonus, the total emitted
power can be dramatically cut and therefore the base station will
generate substantially less electromagnetic interference.  This is
important owing to the increased concerns of electromagnetic exposure.

\item Massive MIMO enables a significant \emph{reduction of latency} on the air interface.

  The performance of wireless communications systems is normally
  limited by \emph{fading}.  The fading can render the received signal
  strength very small at some times. This happens when the signal sent
  from a base station travels through multiple paths before it reaches
  the terminal, and the waves resulting from these multiple
  paths interfere destructively.  It is this fading that makes it hard
  to build low-latency wireless links. If the terminal is trapped in a
  fading dip, it has to wait until the propagation channel has
  sufficiently changed until any data can be received.  Massive MIMO
  relies on the law of large numbers and beamforming in order to avoid
  fading dips, so that fading no longer limits latency.

\item Massive MIMO \emph{simplifies the multiple-access layer}.
 
Owing to the law of large numbers, the channel \emph{hardens} so that
frequency-domain scheduling no longer pays off.  With OFDM, each
subcarrier in a massive MIMO system will have substantially the same
channel gain.  Each terminal can be given the whole bandwidth, which
renders most of the physical-layer control signaling redundant.

\item Massive MIMO increases the \emph{robustness} both to unintended man-made
  interference and to  intentional jamming.  

Intentional jamming of civilian wireless systems is a growing concern
and a serious cyber-security threat that seems to be little known to
the public.  Simple jammers can be bought off the Internet for a few
\$100, and equipment that used to be military-grade can be put
together using off-the-shelf software radio-based platforms for a few
\$1000.  Numerous recent incidents, especially in public safety
applications, illustrate the magnitude of the problem.  During the EU
summit in Gothenburg, Sweden, in 2001, demonstrators used a jammer
located in a nearby apartment and during critical phases of the riots,
the chief commander could not reach any of the engaged 700 police
officers \cite{SPW2013}.
 
Due to the scarcity of bandwidth, spreading information over frequency
just is not feasible so the only way of improving robustness of
wireless communications is to use multiple antennas.  Massive MIMO
offers many excess degrees of freedom that can be used to cancel
signals from intentional jammers.  If massive MIMO is implemented by
using uplink pilots for channel estimation, then smart jammers could
cause harmful interference with modest transmission power.  However,
more clever implementations using joint channel estimation and
decoding should be able to substantially diminish that problem.
  
\end{itemize} 

\section{Limiting Factors of Massive MIMO}
 
 \subsection{Channel Reciprocity}
 
TDD operation relies on channel reciprocity.  There appears to be a
reasonable consensus that the propagation channel itself is
essentially reciprocal, unless the propagation is affected by
materials with strange magnetic properties.  However, the hardware
chains in the base station and terminal transceivers may not be
reciprocal between the uplink and the downlink.  
Calibration of the hardware chains does not seem to constitute a
serious problem and there are calibration-based solutions that have
already been tested to some extent in practice
\cite{Kaltenberger,SYA2012}. Specifically, \cite{SYA2012} treats
reciprocity calibration for a 64-antenna system in some detail and claims a successful
experimental implementation. 

Note that calibration of the terminal uplink and downlink chains is
not required in order to obtain the full beamforming gains of massive
MIMO: if the base station equipment is properly calibrated then the
array will indeed transmit a coherent beam to the terminal. (There
will still be some mismatch within the receiver chain of the terminal
but this can be handled by transmitting pilots through the beam to the
terminal; the overhead for these supplementary pilots is very small.)
Absolute calibration within the array is not required. Instead, as
proposed in \cite{SYA2012}, one of the antennas can be treated as a
reference and signals can be traded between the reference antenna and
each of the other antennas to derive a compensation factor for that
antenna. It may be possible entirely to forgo reciprocity calibration
within the array; for example if the maximum phase difference between
the up-link chain and the down-link chain were less than 60 degrees
then coherent beam forming would still occur (at least with MRT
beamforming) albeit with a possible 3~dB reduction in gain.

\subsection{Pilot Contamination}\label{sec:pilcont}

Ideally every terminal in a Massive MIMO system is assigned an
orthogonal uplink pilot sequence.  However the maximum number of
orthogonal pilot sequences that can exist is upper-bounded by the
duration of the coherence interval divided by the channel
delay-spread. In \cite{Mar2010}, for a typical operating scenario, the
maximum number of orthogonal pilot sequences in a one millisecond
coherence interval is estimated to be about 200. It is easy to exhaust
the available supply of orthogonal pilot sequences in a multi-cellular
system.

The effect of re-using pilots from one cell to another, and the associated
negative consequences, is termed ``pilot contamination''.  More
specifically, when the service-array correlates its received pilot
signal with the pilot sequence associated with a particular terminal
it actually obtains a channel estimate that is contaminated by a
linear combination of channels to the other terminals that share the
same pilot sequence. Downlink
beamforming based on the contaminated channel estimate results in
interference that is directed to those terminals that share the same
pilot sequence. Similar interference is associated with uplink
transmissions of data. This directed interference grows with the
number of service-antennas at the same rate as the desired signal
\cite{Mar2010}. Even partially correlated pilot sequences result in
directed interference.
 
Pilot contamination as a basic phenomenon is not really specific to
massive MIMO, but its effect on massive MIMO appears to be much more profound than in
classical MIMO \cite{Mar2010,HTD2012}.  In \cite{Mar2010} it was
argued that pilot contamination constitutes an \emph{ultimate} limit
on performance, when the number of antennas is increased without bound,
at least with receivers that rely on pilot-based
  channel estimation. While this argument has been contested recently
\cite{ralf2013}, at least under some specific assumptions on the power
control used,  it appears likely that pilot contamination must be
dealt with in some way. This can be done in several ways: 
\begin{itemize}
\item The allocation of pilot waveforms can be optimized. One
  possibility is to use a less-aggressive frequency re-use factor for
  the pilots (but not necessarily for the payload data)---say 3 or
  7. This pushes mutually-contaminating cells farther apart.  It is
  also possible to coordinate the
  use of pilots or adaptively allocate pilot sequences to the
  different terminals in the network \cite{YGF2013}. Currently, the
  optimal strategy is unknown.

\item Clever channel estimation algorithms \cite{ralf2013}, or even blind techniques
  that circumvent the use of pilots altogether \cite{ngo2012evd}, may
  mitigate or eliminate the effects of pilot contamination.  The most
  promising direction seems to be blind techniques that jointly
  estimate the channels and the payload data.

\item New precoding techniques that take into account the network
  structure, such as pilot contamination precoding \cite{AsM2012},
  can utilize cooperative transmission over a multiplicity of
  cells---outside of the beamforming operation---to nullify, at least
  partially, the directed interference that results from pilot
  contamination.  Unlike coordinated beamforming over multiple cells
  which requires estimates of the actual channels between the
  terminals and the service-arrays of the contaminating cells,
  pilot-contamination precoding requires only the corresponding
  slow-fading coefficients. Practical pilot-contamination precoding
  remains to be developed.

\end{itemize}

\subsection{Radio Propagation and Orthogonality of Channel Responses}\label{sec:prop}
  
Massive MIMO (and especially MRC/MRT processing) relies to a large extent on a property of the radio
environment called \emph{favorable propagation}. Simply stated,
favorable propagation means that the propagation channel responses from
the base station to different terminals are sufficiently different. To
study the behavior of massive MIMO systems, channel measurements have
to be performed using realistic antenna arrays. This is so because the
channel behavior using large arrays differs from that usually
experienced using conventional smaller arrays. The most
important differences are that (i) there might be large scale fading over the
array and (ii) the small-scale signal statistics may also change over the
array. Of course, this is also true for physically smaller arrays with
directional antenna elements pointing in various
directions. 

\begin{figure}[t!]
\centerline{
\begin{psfrags}
          \psfrag{u}[cu]{$ 2$}
 \includegraphics[width=14cm]
      {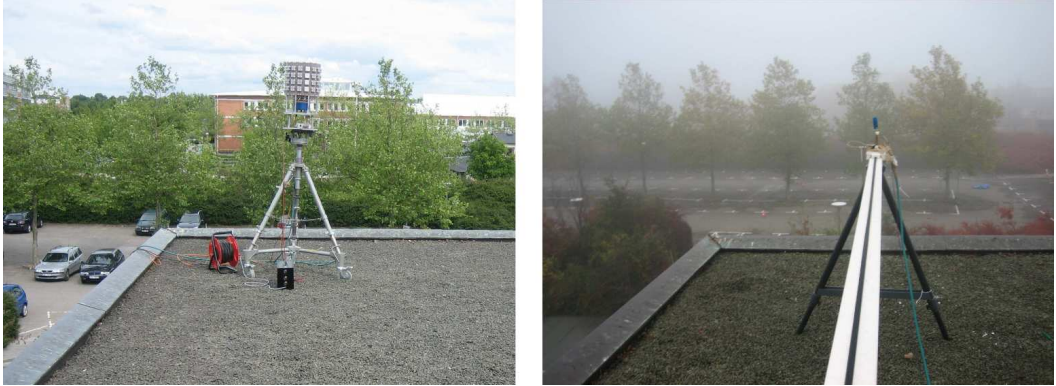}
\end{psfrags}
} \caption{\label{fig:array}Massive MIMO antenna arrays used for the measurements.}
 \end{figure}

Fig.~\ref{fig:array} shows pictures of the two massive MIMO arrays
used for the measurements reported in this paper. To the left is a compact circular massive MIMO array with
128 antenna ports. This array consists of 16 dual-polarized
patch antenna elements arranged in a circle, with 4 such circles
stacked on top of each other. Besides having the advantage of being
compact, this array also provides the possibility to resolve
scatterers at different elevations, but it suffers from worse
resolution in azimuth due to its limited aperture. To the right is a
physically large linear (virtual) array, where a single
omnidirectional antenna element is moved to 128 different positions in an
otherwise static environment to emulate a real array with the same
dimensions. 

\begin{figure}[t!]
\centerline{
 \includegraphics[width=11cm]
      {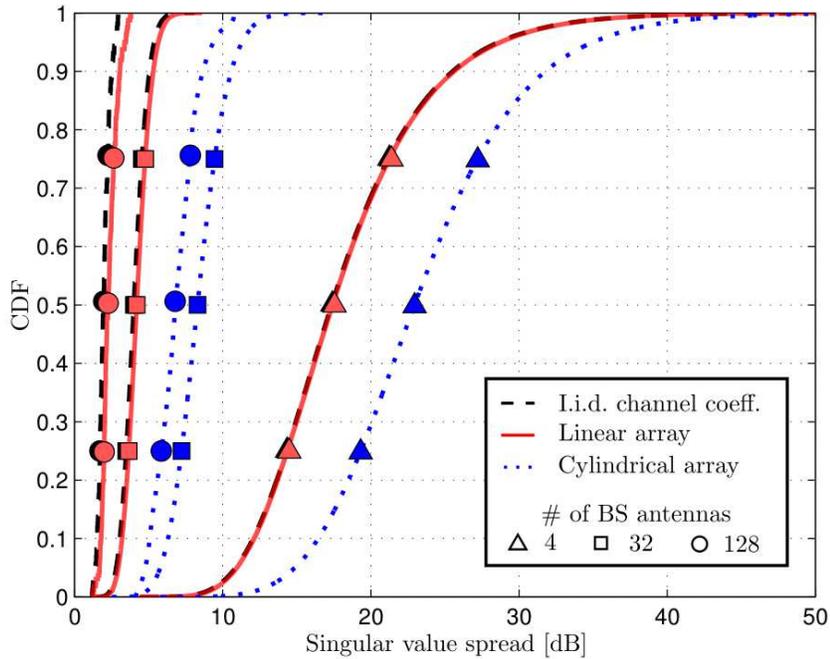}
} \caption{\label{fig:SingValSpread}CDF of the singular value spread 
for MIMO systems with 4 terminals and three different numbers of BS antennas: 4,
32, and 128. The theoretical i.i.d.\ channel is shown as a reference,
while the other two cases are measured channels with linear and
cylindrical array structures at the BS. \emph{Note:} The
curve for the linear array coincides with that of the 
i.i.d.\ channel for 4 BS.}
\end{figure}

One way of quantifying how different the channel responses to
different terminals are, is to look at the spread between the smallest
and largest \emph{singular values} of the matrix that contains the
channel responses. Fig.~\ref{fig:SingValSpread} illustrates this for
a case with 4 user terminals and a base station having 4, 32 and 128
antenna ports, respectively, configured either as a physically large
single-polarized linear array or a compact dual-polarized circular
array.  More specifically, the figure shows the cumulative density
function (CDF) of the difference between the smallest and the largest
singular value for the different measured (narrowband) frequency
points in the different cases. As a reference we
also show simulated results for ideal independent, identically distributed
(i.i.d.) channel matrices, often used in theoretical studies. The
measurements were performed outdoors at the Lund University campus area. The
center frequency was 2.6 GHz an the measurement bandwidth 50 MHz. When using
the cylindrical array, the RUSK Lund channel sounder was employed, while a
network analyzer was used for the synthetic linear array measurements. The
first results from the campaign were presented in \cite{GTER2012}.

For the 4-element array, the median of the singular value spread is
about 23 dB and 18 dB, respectively. This number is a measure of the
fading margin, the additional power that has to be used in order to
serve all users with a reasonable received signal power. With the
massive linear array, the spread is less than 3 dB. In addition, note
that none of the curves has any substantial tail. This means that the
probability of seeing a singular value spread larger than 3 dB,
anywhere over the measured bandwidth, is essentially negligible.

\begin{figure}[t!]
\centerline{ \includegraphics[width=11cm] {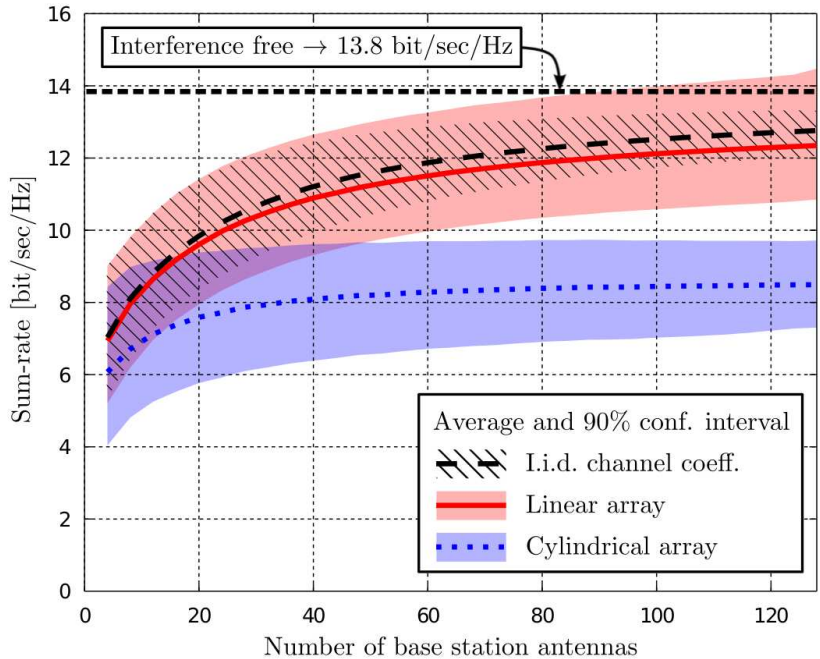}
} \caption{\label{fig:MRTsumrate} Achieved downlink sum-rates, using MRT
precoding, with four single-antenna terminals and between 4 and 128 base station
  antennas.}
\end{figure}

To further illustrate the influence of different number of antenna
elements at the base station and the antenna configuration, we plot in
Fig.~\ref{fig:MRTsumrate} the sum-rate for 4 closely spaced users
(less than 2 meters between each user at a distance of about 40 m from
the base station) in a non line-of-sight (NLOS) scenario when using
MRT as pre-coding. The transmit power is normalized so
that on the average, the interference free signal-to-noise-ratio at
the terminals is 10 dB.

As can be seen in Fig.~\ref{fig:MRTsumrate}, the sum-rate approaches
that of the theoretical interference-free case as the number of
antennas at the base station increases. The shaded areas in red (for
the linear array) and blue (for the circular array) shows the 90 percent
confidence levels of the sum-rates for the different narrowband frequency
realizations. As before, the variance of the sum-rate decreases as the
number of antennas increases, but slowly for the measured channels. The slow
decrease can, at least partially, be attributed to the shadow fading occurring
across the arrays. For the linear array in the form of shadowing by external
objects along the array and for the cylindical array as shadowing caused by
directive antenna elements pointing in the wrong direction.
The performance of the physically large array approaches that of the theoretical
i.i.d.\ case when the number of antennas grows large. The compact circular array
has inferior performancev compared to the linear array owing to its smaller aperture --
it cannot resolve the scatterers as well as the physically large array -- and its directive
antenna elements sometimes pointing in the wrong direction.
Also, due to the fact that most of the scatterers are seen at the same horizontal angle,
the possibility to resolve scatters at different elevations
gives only marginal contributions to the sum-rate in this scenario. 

It should be mentioned here that when using somewhat more complex, but
still linear, pre-coding methods such as zero-forcing or minimum mean-square 
error, the convergence to the i.i.d.\ channel performance is faster and
the variance of the sum-rate is lower as the number of base station
antennas is increased, see \cite{GTER2012} for further
details. Another aspect worth mentioning is that also for a very
tricky propagation scenario, such as closely spaced users in
line-of-sight conditions, it seems that the large array is able to
separate the users to a reasonable extent using the different spatial
signatures that the users have at the base station due to the enhanced spatial
resolution.  This would not be possible with conventional MIMO.  Those
conclusions are also in line with the observations in
\cite{HHW2012} where another outdoor measurement campaign
is described and analyzed.

Overall, there is compelling evidence that the assumptions on
favorable propagation that underpin massive MIMO are substantially
valid in practice. Depending on the exact configuration of the large
array and the pre-coding algorithms used the convergence towards the
ideal performance may be faster or slower, as the number of antennas
is increased. However, with about 10 times more base station antennas
 than the number of users, it seems that it is possible to get
a stable performance not far from the theoretically ideal performance
also under what is normally considered very difficult propagation conditions.
 
\section{Massive MIMO: a Goldmine of Research Problems}

While massive MIMO renders many traditional problems in communication
theory less relevant, it uncovers entirely new problems that need
research:
\begin{itemize}

\item \emph{Fast and distributed, coherent signal processing.}
  Massive MIMO arrays generate vast amounts of baseband data that need
  be processed in real time.  This processing will have to be simple,
  and simple means linear or nearly linear.  Fundamentally, this is
  good in many cases, cf.\ Fig.~\ref{mrc}.  Much research needs be
  invested into the design of optimized algorithms and their
  implementation.  On the downlink, there is enormous potential for
  ingenious precoding schemes. Some examples of recent work in this
  direction include \cite{ZYP2013}.

 \item \emph{The challenge of low-cost hardware}. Building hundreds of
   RF chains, up/down converters, A/D--D/A converters, and so forth,
   will require economy of scale in manufacturing comparable to what
   we have seen for mobile handsets.

\item \emph{Hardware impairments.} Massive MIMO relies on the law of
  large numbers to average out noise, fading and to some extent,
  interference.  In reality, massive MIMO must be built with low-cost
  components. This is likely to mean that hardware imperfections are
  larger: in particular, phase noise and I/Q imbalance.  Low-cost and
  power-efficient A/D converters yield higher levels of quantization
  noise. Power amplifiers with very relaxed linearity requirements
  will necessitate the use of per-antenna low-peak-to-average
  signaling, which as already noted, is feasible with a large excess
  of transmitter antennas.  With low-cost phase locked-loops or even
  free-running oscillators at each antenna, phase noise may become a
  limiting factor. However, what ultimately matters is how much the
  phase will drift between the point in time when a pilot symbol is
  received and the point in time when a data symbol is received at
  each antenna.  There is great potential to get around the phase
  noise problem by design of smart transmission physical-layer schemes
  and receiver algorithms.

 \item \emph{Internal power consumption.}  Massive MIMO offers the
   potential to reduce the \emph{radiated} power a thousand times, and
   at the same time drastically scale up data rates. But in practice,
   the \emph{total} power consumed must be considered, that includes
   the cost of baseband signal processing.  Much research must be
   invested into highly parallel, perhaps dedicated, hardware for the
   baseband signal processing.
 
 \item \emph{Channel characterization.} There are additional
   properties of the channel to consider when using massive MIMO
   instead of conventional MIMO. To facilitate a realistic performance
   assessment of massive MIMO systems it is necessary to have channel
   models that reflect the true behavior of the radio channel,
   i.e. the propagation channel including effects of realistic antenna
   arrangements. It is also important to develop more sophisticated
   analytical channel models.  Such models need not necessarily be
   correct in every fine detail, but they must capture the essential
   behavior of the channel.  For example, in conventional MIMO the
   Kronecker model is widely used to model channel correlation---this
   model is not an exact representation of reality but provides a useful
   model for certain types of analysis, despite its limitations. A similar way of
   thinking could probably be adopted for massive MIMO channel
   modeling.

\item \emph{Cost of reciprocity calibration.} TDD will require
  reciprocity calibration. How often must this be done and what is the
  best way of doing it? What is the cost, in terms of time- and
  frequency resources needed to do the calibration, and in terms of
  additional hardware components needed?

\item \emph{Pilot contamination.}  It is likely that pilot
  contamination imposes much more severe limitations on massive MIMO
  than on traditional MIMO systems.  We discussed some of the issues
  in detail, and outlined some of the most relevant research
  directions in Section~\ref{sec:pilcont}.
 
 \item \emph{Non-CSI@TX operation.}  Before a link has been
   established with a terminal, the base station has no way of knowing
   the channel response to the terminal.  This means that no array
   beamforming gain can be harnessed.  In this case, probably some
   form of space-time block coding is optimal. Once the terminal has
   been contacted and sent a pilot, the base station can learn the
   channel response and operate in coherent MU-MIMO beamforming mode,
   reaping the power gains offered by having a very large array.
 
 \item \emph{New deployment scenarios}.  It is considered
   extraordinarily difficult to introduce a radical new wireless
   standard. One possibility is to introduce dedicated applications of
   massive MIMO technology that do not require backwards
   compatibility. For example, as discussed in
   Section~\ref{sec:potential}, in rural areas, a billboard-sized
   array could provide 20 Mbit/s service to each of a thousand homes
   using special equipment that would be used solely for this
   application.  Alternatively, a massive array could provide the
   backhaul for base stations that serve small cells in a densely
   populated area.  So rather than thinking of massive MIMO as a
   competitor to LTE, it can be an enabler for something that was just
   never before considered possible with wireless technology.

\item \emph{System studies and relation to small-cell and HetNet
  solutions}. The driving motivation of massive MIMO is to
  simultaneously and drastically increase the data rates and the
  overall energy efficiency.  Other potential ways of reaching this
  goal are network densification by the deployment of small cells,
  resulting in a heterogeneous architecture, or coordination of the
  transmission of multiple individual base stations. From a pure
  fundamental perspective, the ultimately limiting factor of the
  performance of any wireless network appears to be the availability
  of good enough channel state information (CSI), that facilitates
  phase-coherent processing at multiple antennas or multiple access
  points \cite{lozano2012fundamental}.  Considering factors like
  mobility, Doppler shifts, phase noise and clock synchronization,
  acquiring high-quality CSI seems to be easier with a collocated
  massive array then in a system where the antennas are distributed
  over a large geographical area.  But at the same time, a distributed
  array or small-cell solution may offer substantial path-loss gains
  and would also provide some diversity against shadow fading.  The
  deployment costs of a massive MIMO array and a distributed or
  small-cell system also are likely to be very different. Hence, both
  communication-theoretic and techno-economic studies are needed to
  conclusively determine which approach that is superior. However, it
  is likely that the winning solution will comprise a combination of
  all available technologies.

\item \emph{Prototype development.}  While massive MIMO is in its
  infancy, basic prototyping work of various aspects of the technology
  is going on in different parts of the world. The Argos testbed
  \cite{SYA2012} was developed at Rice university in cooperation with
  Alcatel-Lucent, and shows the basic feasibility of the massive MIMO
  concept using 64 coherently operating antennas.  In particular, the
  testbed shows that TDD operation relying on channel reciprocity is possible.
  One of the virtues of the Argos testbed in particular is that it is
  entirely modular and scalable and that is built around commercially
  available hardware (the WARP platform).  Other test systems around
  the world also have demonstrated the basic feasibility of scaling up
  the number of antennas. The Ngara testbed in Australia
  \cite{SuzukiKAGHPBMC2012} uses a 32-element base station array to
  serve up to 18 users simultaneously with true spatial multiplexing.
  Continued testbed development is highly desired both to prove the
  massive MIMO concept with even larger numbers of antennas, and to
  discover potentially new issues that need urgent research.
 \end{itemize}
 
\section{Conclusions and Outlook}

In this paper we have highlighted the large potential of massive MIMO
systems as a key enabling technology for future beyond 4G cellular
systems. The technology offers huge advantages in terms of energy
efficiency, spectral efficiency, robustness and reliability. It allows
for the use of low-cost hardware both at the base station as well as
at the mobile unit side. At the base station the use of expensive and
powerful, but power-inefficient, hardware is replaced by massive use
of parallel low-cost, low-power units that operate coherently
together. There are still challenges ahead to realize the full
potential of the technology, e.g., when it comes to computational
complexity, realization of distributed processing algorithms, and
synchronization of the antenna units. This gives researchers both in
academia and industry a goldmine of entirely new research problems to
tackle.

\section*{Acknowledgement}
The authors would like to thank Xiang Gao, doctoral student at Lund
University, for her analysis of channel measurements presented in
figures \ref{fig:SingValSpread}--\ref{fig:MRTsumrate}, and the Swedish
organizations ELLIIT, VR, and SSF, for their funding of parts of this work.

% \bibliographystyle{IEEEtran}
% \bibliography{IEEEabrv,ComMag}


\newpage
\section*{Biographies}

\textbf{Erik G. Larsson}  is Professor and Head of the Division
for Communication Systems in the Department of Electrical Engineering
(ISY) at Link\"oping University (LiU) in Link\"oping, Sweden.
He has published some 100 journal papers
on signal processing and communications and he is co-author of the textbook \emph{Space-Time
Block Coding for Wireless Communications}. He
is Associate Editor for the \emph{IEEE Transactions on 
Communications} and he
received the \emph{IEEE Signal Processing Magazine} Best Column Award 2012.

\textbf{Ove Edfors} is Professor of Radio Systems at the Department
of Electrical and Information Technology, Lund University, Sweden.
His research interests include statistical signal
processing and low-complexity algorithms with applications
in wireless communications. In the context of Massive MIMO, his
research focus is on how realistic propagation characteristics 
influence system performance and base-band processing complexity.

\textbf{Fredrik Tufvesson} received his Ph.D. in 2000 from Lund University
in Sweden. After almost two years at a startup company, Fiberless Society,
he is now associate professor at the department of Electrical and Information
Technology, Lund University. His main research interests are channel
measurements and modeling for wireless communication, including channels
for both MIMO and UWB systems. Beside this, he also works on distributed
antenna systems and radio based positioning.

\textbf{Thomas L.~Marzetta} received the PhD in electrical engineering
from MIT. He joined Bell Laboratories in 1995. Within the former Mathematical
Sciences Research Center he was director of the Communications and Statistical
Sciences Department. He was an early proponent of Massive MIMO which can provide
huge improvements in wireless spectral-efficiency and energy-efficiency over 4G
technologies. He received the 1981 ASSP Paper Award, became a Fellow of the IEEE
in Jan.~2003, and received the 2013 IEEE Guglielmo Marconi Best Paper Award.

 \end{document}